\documentclass[twocolumn,prd,amssymb,amsmath,preprintnumbers,floatfix,aps,nofootinbib]{revtex4-1}
\pdfoutput=1
\usepackage{dcolumn,graphicx,bm,psfrag}
\usepackage[colorlinks=true,citecolor=blue,urlcolor=black,linktocpage=true,
linkcolor=blue]{hyperref}

\newcommand{\be}{\begin{equation}}
\newcommand{\ee}{\end{equation}}
\newcommand{\bear}{\begin{eqnarray}}
\newcommand{\eear}{\end{eqnarray}}
\newcommand{\ba}{\begin{array}}
\newcommand{\ea}{\end{array}}

\begin{document}
\title{\boldmath \bf \Large  Hunting sub-GeV dark matter with NO$\nu$A  near detector} 
\author{\bf Patrick deNiverville} 
\affiliation{Center for Theoretical Physics of the Universe, IBS, Daejeon 34126, Korea}
\author{\bf Claudia Frugiuele}
\affiliation{Department of Particle Physics and Astrophysics, Weizmann Institute of Science, Rehovot, Israel}

\begin{abstract}

We study the sensitivity of the NO$\nu$A near detector to MeV-GeV dark matter while operating symbiotically with the neutrino program. 
We find that NO$\nu$A  could explore a large new region of parameter space over the next few years for dark matter masses below 100 MeV, reaching the thermal target for a scalar dark matter particle for some masses.
This result represents a significant improvement over existing probes such as Babar, E137, and LSND.
\end{abstract}

\maketitle

\section{Introduction}
\label{sec:intro}
There are many compelling dark matter (DM) candidates with a correspondingly wide range of possible masses and couplings to the visible sector. Probing this vast parameter space requires a correspondingly broad experimental program, and neutrino fixed target facilities can play a role in this quest by searching for signatures of DM scattering with electrons and/or nuclei in their (near) detectors~\cite{Batell:2009di, deNiverville:2011it, deNiverville:2012ij, Dharmapalan:2012xp, Batell:2014yra, Soper:2014ska,Dobrescu:2014ita,Coloma:2015pih,Frugiuele:2017zvx,millicharged}. Their main advantage lies in the high luminosity available, frequently boasting $ 10^{20}-10^{21}$ protons on target (POT) per year, which allows for the production of a sizeable relativistic DM beam.
Moreover, this setup offers the possibility of probing light DM/quark couplings, complementary to direct detection experiments sensitive to electron/DM interactions \cite{Essig:2012yx,cosmicvision}.

However, the neutrino background presents a significant challenge when searching for nucleon-DM scattering \cite{deNiverville:2011it,Dharmapalan:2012xp,Batell:2014yra,Coloma:2015pih,Dobrescu:2014ita,Frugiuele:2017zvx}. More promising in this regard is electron-DM scattering, where the neutrino related backgrounds are much smaller. Some of the strongest constraints on the DM parameter space have been placed by recasting existing neutrino-electron scattering data from the LSND experiment \cite{lsnde,lsndpatrick} and we aim to investigate whether present neutrino facilities could improve on LSND's sensitivity. In particular, we study the reach of the NO$\nu$A near detector to DM-electron scattering.
By reinterpreting an existing analysis on $\nu-e$ elastic scattering we find sensitivity to a large region of the DM parameter space still unconstrained by present experimental probes such as LSND \cite{ lsnde,lsndpatrick}, E137 \cite{Batell:2014mga}, NA64 \cite{Banerjee:2017hhz}, BaBar \cite{Lees:2017lec} and CRESST-II \cite{Angloher:2015ewa}. 

The paper is organized as follow: in Sec. \ref{sec:Vec_Portal} we define our benchmark model. Sec. \ref{sec:DM_prod} summarises the main aspects of DM searches at neutrino facilities. In Sec. \ref{sec:scatter}, we present the sensitivity of NOV$\nu$A of electron-DM elastic scattering by recasting of current analysis performed by the collaborations. Finally, we present a summary in Sec. \ref{sec:conc}.
 \section{ Vector Portal}
 \label{sec:Vec_Portal}
Our benchmark model consists of a dark photon (DP) \cite{Holdom:1985ag}  $A'_{\mu}$, the gauge boson of a new dark gauge group $U(1)_D$ kinetically mixed with the photon, and a scalar $\chi$ charged under $U(1)_D$ that serves as a DM candidate:
\be 
\mathcal{L}_{ \rm DM}=\mathcal{L}_{A^{\prime}}+\mathcal{L}_\chi 
\ee
where:
\be
\mathcal{L}_{A^{\prime}} =- \frac{1}{4} F'_{\mu \nu}F^{\prime \mu \nu} +\frac{m^2_{A'} }{2}A^{\prime \mu} A^{\prime}_{ \mu}-\frac{1}{2} \epsilon \;  F^{\prime}_{\mu \nu} F^{\mu \nu},
\ee
where $  \epsilon $ is the DP-photon kinetic mixing, while:
 \be
\mathcal{L}_\chi =  \frac{i g_D}{2}  A^{\prime \mu} J_{\mu}^{\chi}+\frac{1}{2}  \partial_{\mu} \chi^\dagger \partial^{\mu} \chi - m_{\chi}^2 \chi^\dagger \chi,
\ee
where $ J_{\mu}^{\chi}=  \left[ (\partial_\mu \chi^\dagger) \chi  -  \chi^\dagger  \partial_\mu \chi \right]$ and  $g_D$ is the $U(1)_D$ gauge coupling.
The region of the parameter space reachable by neutrino facilities is $ m_{A'} > 2 m_{\chi}$ and $ g_D \gg \epsilon e$ which implies that the DP almost always decays into a $\chi\chi^\dagger$ pair. 

For much of the parameter space studied, the strongest experimental constraints for $ m_{\chi} >60$\,MeV  come from a monophoton search performed by BaBar \cite{Lees:2017lec} that excludes the existence of a DP with $ \epsilon > 10^{-3}$ and $ m_{A'} < 8$ GeV decaying into $\chi\bar\chi$. For large values of $\alpha_D$, CRESST-II places strong constraints on $m_\chi > 500$\,MeV. The NA64 collaboration has recently published very strong limits for DP masses below 100\,MeV \cite{Banerjee:2017hhz} via a missing energy analysis. However, for large $ \alpha_D$, NA64 sensitivity is superseded by experiments looking at electron-DM scattering such as LSND \cite{lsnde,lsndpatrick}, and E137 \cite{Batell:2014mga}.
These constraints do not depend on whether the particle $\chi$ produced through prompt DP decay is DM or not, as the only necessary ingredient is its stability with respect to the target-detector distance (a few kilometers at most).

We are particularly interested in the region where $\chi$ is a thermal relic compatible with the observed dark matter relic energy density. A complex scalar dark matter candidate $\chi$ is safe from constraints coming from precise measurements of the temperature anisotropies of the cosmic microwave background (CMB) radiation \cite{Lin:2011gj,Ade:2015xua}. Other choices for DM  not in tension with the CMB are a Majorana or Pseudo-Dirac fermion. Furthermore, for the minimal DP model considered here a complex scalar lighter than 6.9 MeV is ruled out \cite{Boehm:2013jpa} by the Planck measurement of $N_{eff}$ \cite{Ade:2015xua}. 
 
For $ m_{A'} > 2 m_{\chi}$, the annihilation cross section for a scalar dark matter particle can be written as \cite{Gordan}:
\be
\sigma (\chi \chi \to f \bar f) v  \sim \frac{8  \pi v^2 Y } { m_{\chi}^2}, 
\label{xthermal}
\ee
where $v$ is the relative DM  velocity and $Y$ is defined as:
\be
Y \equiv \epsilon^2 \alpha_D \left (\frac{ m_{\chi}}{m_{A'}}\right)^4 ;
\ee
in the following we will present the sensitivity of neutrino facilities in the $ (Y,m_{\chi})$ plane by considering as benchmark points $ \alpha_D=0.5$ and $\alpha_D = 0.05$ with 
$m_{A'}= 3 m_{\chi}$. This choice has the advantage of making the so called thermal targets apparent, that is regions of the parameter space where, for a certain scenario, the correct thermal abundance is obtained \cite{Gordan,cosmicvision}. For a complex scalar and for our benchmark point, BaBar bounds constrain thermal relics to be lighter than 500 MeV \cite{Lees:2017lec}. Hence, the unexplored parameter space for which $\chi$ provides a good thermal relic dark matter candidate is large in this simple scenario. 
The $U(1)$ gauge coupling $\alpha_D$ is bounded by the constraint on DM self-scattering cross-section coming on halo shape and bullet clusters observations, that is
\be
\frac{\sigma}{m_{\chi}} \lesssim \rm few \times cm^{2}/g,
\ee
which however does not lead to a significant bound in the region $m_{A'} > m_{\chi}$ \cite{Izaguirre:2013uxa}. We will limit $\alpha_D \lesssim 0.5$, where this upper bound is suggested by the running of $\alpha_D$ \cite{Davoudiasl:2015hxa}.

\section{DM production and detection at neutrino facilities}
\label{sec:DM_prod}

The near detector (ND) of a fixed target neutrino facility, designed to measure the neutrino flux before significant oscillations occur, also offers the opportunity for measurement of neutrino interactions. As pointed out in Ref. \cite{Batell:2009di}, the ND could also serve as a DM detector. The idea is as follows: the DP is produced in the interaction of the proton beam with the target and it then decays promptly into DM particles, producing a DM beam alongside the neutrino beam. The DM beam is then detected inside of the ND through DM-nucleon or DM-electron interactions.
DPs are produced mainly by rare decays of $ \pi^0$ and $\eta$, or for heavier masses, via proton bremsstrahlung and mixing between the DP and the $\rho$.

We simulate the DP production $p N \to A'X $  via the BdNMC simulation tool \cite{deNiverville:2016rqh} which considers only the primary reactions resulting from the collision of the proton beam with the target. This is sufficient to capture the leading order production mode for the DP, but does miss possible secondary and tertiary production from showering within the target. As the DM produced in the decays of the secondary or tertiary DP have much lower energy and a far larger angular spread than those produced by the initial proton-target collisions, their contribution to the final dark matter event rate is highly suppressed. A more thorough analysis would also consider these additional contributions to the DM beam, and our projections should be considered conservative.

\begin{table}[h]
        
\begin{center}
  \begin{tabular}{ | l  | c | c|  c | c|r|}
    \hline
    Detector  & $d (m)$ & $L_d$ (m) & $\theta (mrad)$ & $n_e( 1/m^3)$ \\ \hline
   NO$\nu$A & 990 & 14.3 & 14.6 & $4.13\times10^{29}$ \\ \hline
     \end{tabular}

    \caption{Main specifications of NO$\nu$A near detector.}
      
    \label{tab:specs}
 \end{center}
 \end{table}
The total number of DM particles produced in the target via bremsstrahlung is:
\be
N_{\chi }= \frac{ 2 N_{\rm POT}} {\sigma_{\rm T} (pp)} \sigma_{\rm T} ( p p \to A' X)
\label{DMtot}
\ee
where the factor of two takes into account the production of the  $\chi \bar \chi$ pair, $ N_{\rm POT}$ is the number of proton on target, and $ \sigma(pp) \sim 40 $\,mb is the total proton-proton cross section, for a beam of 120\,GeV.
The total number of DM produced through the decay of some pseudoscalar meson $\phi$ is given by:
\begin{equation}
 N_{\chi} = 2 N_\mathrm{POT} N_{\phi\mathrm{/POT}} \mathrm{Br}(\phi \to \chi \chi^\dagger)
\end{equation}
where we take the number of $\phi$'s produced per POT $N_{\phi\mathrm{/POT}}$ to be approximately 1 for $\pi^0$ and $1/30$ for $\eta$. This is an underestimate, as many more mesons are produced in secondary and tertiary beam-target interactions but as previously mentioned, these have smaller energies and a larger angular spread, greatly decreasing their intersection with the detector and suppressing their contribution to the dark matter-electron scattering rate.

Once the DM beam is produced, a fraction $ \epsilon_{\rm det}  \, N_{\chi } $ enters the neutrino (near) detector and scatters with electrons and nucleons.
For $m_{A'} \ll \Lambda_{\rm QCD}$, the DM-electron scattering is the dominant process, while for neutrinos the electron scattering cross section is at least three orders of magnitude suppressed compared to the hadronic cross section \cite{Formaggio:2013kya}, hence this is a powerful signature to explore in the light mass region.

The inclusive electron-neutrino scattering cross section can be approximated by \cite{Formaggio:2013kya}:
\be
\sigma(\nu_l e ) \sim 10^{-42} \left(\frac{E_{\nu}}{\rm GeV}\right) \rm cm^{-2}
\ee
while for $E_{\chi} \gg m_V$ the DM electron elastic cross section is:
\be
\sigma( \chi e) \sim \frac{4 \pi \alpha_D \alpha \epsilon^2 }{m_{A'}^2} \sim  10^{-27} \alpha_D \epsilon^2 \left (\frac{100 \; \rm MeV}{ m_{A'}} \right)^2  \rm cm^{-2}
 \ee 
 such that for $\epsilon \sim 10^{-4}-10^{-5}$ and a light DP the DM-electron scattering cross section is still orders of magnitude larger than the neutrino-electron cross section.
The number of signal events $S_{\chi e \to \chi e } $  is then:
\begin{equation}
S_{\chi e \to \chi e}  =  L_d n_{e} \; \int
dN_{\rm T}  (E_\chi)   \, \sigma( \chi e)  ~~.
\label{DIS1}
\end{equation}
where $n_{e}$ is the detector electron density , while
\begin{equation}
dN_{\rm T}  (E_\chi)  =  \, \epsilon_{\rm det}  \, N_{\chi } \,  \left(\frac{1}{\sigma} \frac{d\sigma}{ dE_\chi}\right)\!\! \left(pN\to \chi\bar{\chi}\right)_{\rm T} \; dE_\chi   ~~.
\label{fractionE}
\end{equation}
where $\epsilon_{\rm det}$ us the acceptance of the detector under investigation.

We now consider the Fermilab NuMI facility \cite{Anderson:1998zza}, which operates with access to the Main Injector's 120 GeV proton beam and delivers neutrinos to several nearby detectors: MINOS, NO$\nu$A, and Minerva.
We focus on the NO$\nu$A (NuMI Off-axis $\nu_e$ Appearance) near detector, and explore its sensitivity to DM masses in the MeV-GeV window produced via the prompt decay of a sub-GeV DP (see \cite{Dobrescu:2014ita,Coloma:2015pih} for a discussion on the sensitivity of these facilities to few-GeV vector bosons). The NO$\nu$A ND is located 990 m away from the target and is 12 milliradians off-axis, where its location was chosen to guarantee that the energy distribution of the neutrino flux peaks at 2 GeV.
The NO$\nu$A near detector is a 300-ton low-Z, highly active tracking calorimeter able to differentiate between muons (long tracks), electrons (EM showers) and $\pi^0$'s (which leave a gap before decaying to $\gamma$'s).
This low-Z material and fine cell structure provides the NO$\nu$A detector  with good angular resolution for neutrino(DM)-electron elastic scattering.
Tab. \ref{tab:specs} presents the specifications of the NO$\nu$A ND, while table Tab. \ref{tab:accept} shows the geometrical acceptance of the primary production modes.

\par
\begin{table}
\begin{center}
  \begin{tabular}{ | l | c | r |}
    \hline
    $A'$ production mode  & $\epsilon_{NO\nu A}$ \\ \hline
   $ \pi^0 \to \gamma A'$ & 0.0027 \\ \hline
    $ p p \to A' p p $  & 0.0043  \\ 
    \hline
    \end{tabular}
    \caption{Geometrical acceptance of the NO$\nu$A near detector considering DM production via pion decay and proton bremsstrahlung.}
    \label{tab:accept}
  \end{center}
 \end{table}

\section{ Electron-dark matter scattering signature in NO$\nu$A near detector}
\label{sec:scatter}
\begin{figure}[h]
 \begin{center}
  \includegraphics[width=0.36\textwidth]{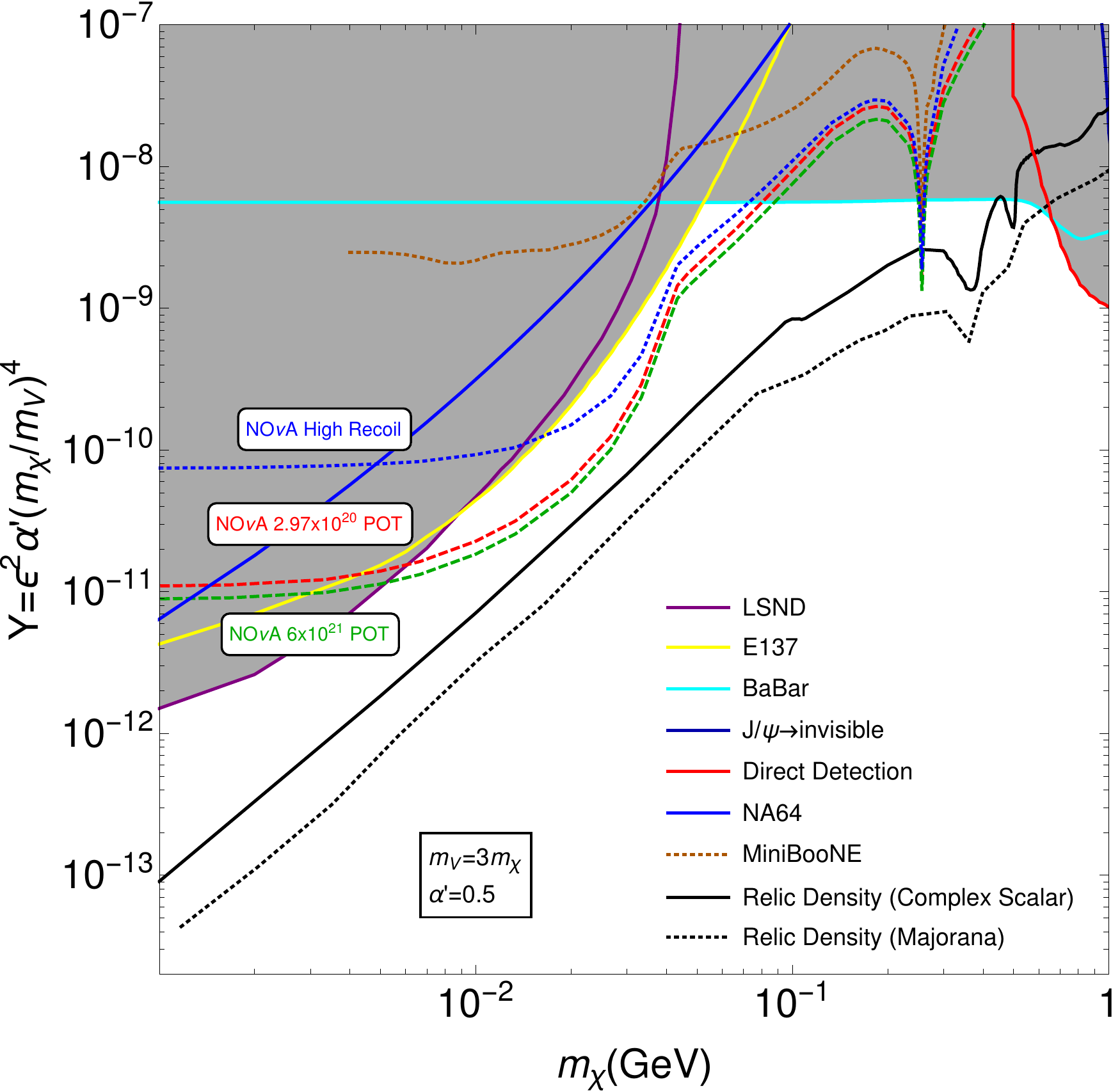}
 \end{center}
     \caption{NO$\nu$A estimated sensitivity to a DP decaying into $\chi \chi^\dagger$ pairs for the benchmark point $\alpha_D =0.5 $ and $ m_{A'}= 3 m_{\chi}$. We plot contours of 58 (753) events for $2.97\times10^{20}$ ($6\times10^{21}$) POT with $E_e\theta^2 < 0.005\, \mathrm{GeV}\,\mathrm{rad}^2$ and for the contour labelled High Recoil, 345 events for $E_e \in [5,15]\,\mathrm{GeV}$ for $6\times10^{21}$. Also shown are the strongest existing experimental limits: LSND \cite{ lsnde,lsndpatrick}, E137 \cite{Batell:2014mga}, NA64 \cite{Banerjee:2017hhz}, BaBar \cite{Lees:2017lec} and CRESST-II \cite{Angloher:2015ewa}.}
\label{escattering1}
\end{figure}
\begin{figure}[h]
 \begin{center}
  \includegraphics[width=0.36\textwidth]{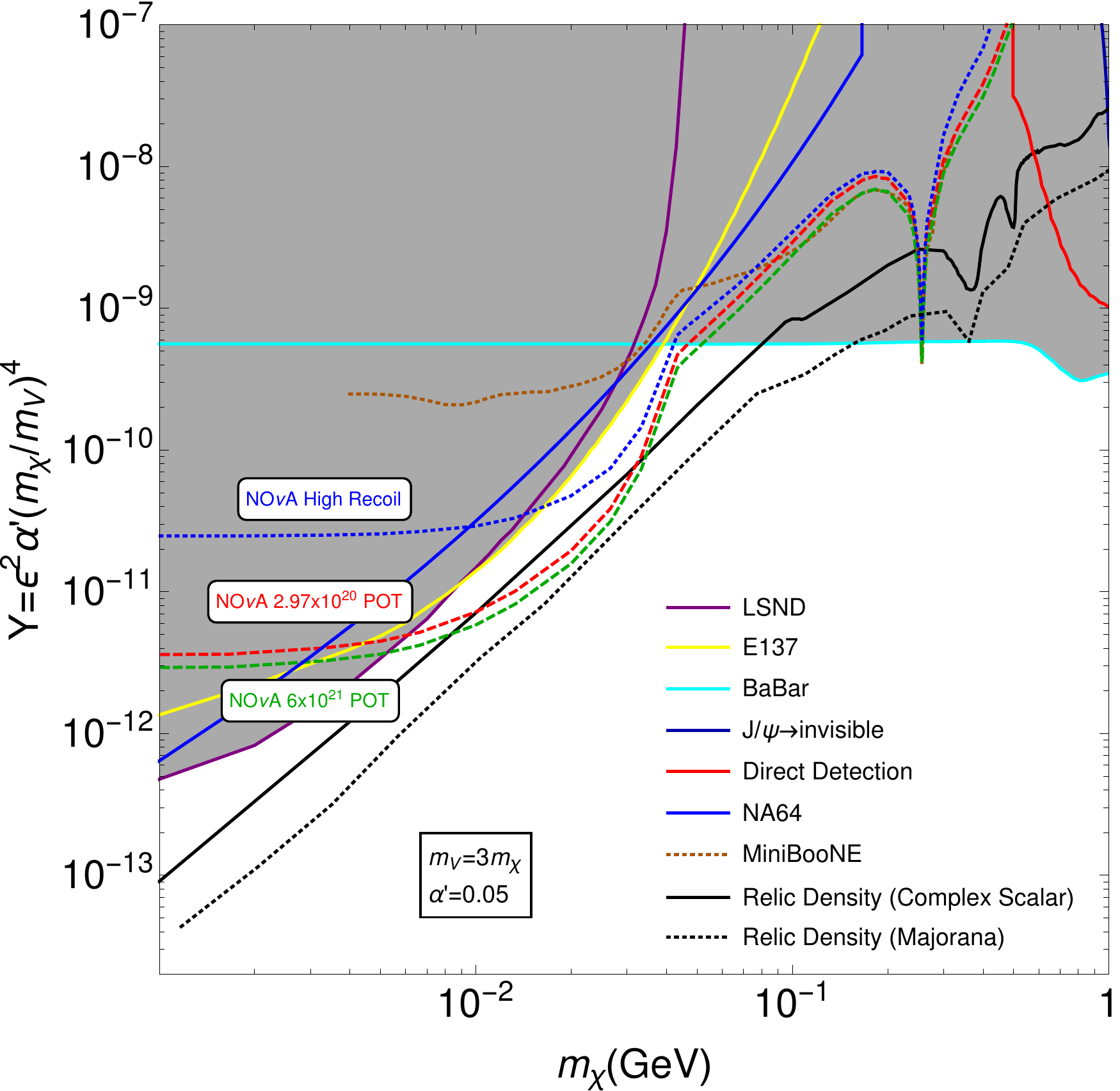}
 \end{center}
     \caption{ NO$\nu$A estimated sensitivity to a DP decaying into electrons for the benchmark point $\alpha_D =0.05 $ and $ m_{A'}= 3 m_{\chi}$. See Fig. \ref{escattering1} for further details.
     }
\label{escattering2}
\end{figure}

We now study NO$\nu$A's sensitivity to electron-DM scattering events by reinterpreting the $\nu-$e elastic scattering analysis \cite{Bian:2017axs} performed on $2.97 \times 10^{20}$ POT as a DM-electron scattering analysis, treating both the $\nu$ signal and background as background for the DM search. This amounts to 160 events: the estimated number of $\nu$-e  signal event is $\sim 140$, while its background of $\sim 20$ \cite{Bian:2017axs} events includes both charged current quasi-elastic (CCQE) events and neutral current (NC) events emitting a single pion.
Each signal event is composed of single forward-boosted electron with energy $E_e$ in the range 0.5 GeV-5 GeV; for these energies the electron detection efficiency is approximately $50 \%$. Since electrons coming from $\nu$-e elastic scattering are very forward along the direction of the neutrino beam we impose a cut on $ E_e \theta^2 <0.005 $.

We then simulated DM production and scattering events using the BdNMC code \cite{deNiverville:2016rqh} and applied the analysis cuts, finding that nearly all dark matter events passed the cuts.
A comment is in order: a $\nu$-electron elastic scattering analysis can be used to measure the neutrino flux and improve the total uncertainty on flux estimates over those based on hadronic measurements (see for instance \cite{Park:2015eqa} for how this is used to measure the Minerva flux). However, the problem of whether this the accuracy of this technique could be adversely affected by DM contamination is an important one which we will address in a separate publication. In the following we will not rely on any flux estimates using neutrion-electron scattering, and instead assume a standard $10\%$ uncertainty in the flux from hadronic calculations This is a conservative choice, and there is a strong effort in improving on that. 
Another strategy suggested in \cite{biao} is to instead consider high energy events with electron recoil energies of 5 GeV-15 GeV without imposing any angular cut. In this case the neutrino background consists of 67 events for $2.97\times10^{20}$ POT \cite{biao} and we will assume a reconstruction efficiency of $20\%$.

Figs. \ref{escattering1} and \ref{escattering2}  illustrate NO$\nu$A  sensitivity for two different benchmark values of $\alpha_D $ (0.5 and 0.05) obtained by recasting the analysis of Ref. \cite{Bian:2017axs} using the angular cut of $ E_e \theta^2 <0.005 $ with $2.97 \times 10^{20}$ POT (labelled NO$\nu$A $2.97 \times 10^{20}$) and the final POT of $6 \times 10^{21}$ (labelled NO$\nu$A $6 \times 10^{21}$). We also considered high energy events with electron recoils between 5 GeV and 15 GeV with $6\times 10^{21}$ POT only (labelled NO$\nu$A High Recoil), though this produced somewhat weaker limits.
NO$\nu$A is able to reach new regions of parameter space above and beyond that of currently existing limits for $ m_{\chi} \lesssim 200$ MeV. The black solid line represents the region where the correct relic abundance is achieved via thermal freeze out into SM particles and NO$\nu$A can probe it in some regions of the parameter space for $\chi$ masses below 100 MeV for $ \alpha_D = 0.05$.
   
Both the LSND and E137 limits were obtained as a recast of existing searches, while in this region the strongest bound from an analysis by an experimental collaboration comes from NA64.
While future MiniBooNE analyses may soon realize limits comparable to LSND and E137, NO$\nu$A should be able to place similar limits with currently available data and improve upon them with $6\times10^{21}$ POT.
At low masses, $ m_{\chi} \lesssim 6$ MeV, NO$\nu$A's reach flattens out because $ m_{A'} \ll E_f$, where $E_f$ is the energy of the dark matter particles after DM-electron scattering, and limits from LSND dominate due to the lower energy threshold. In order to improve NO$\nu$A sensitivity in this region it would be necessary to reduce the threshold for the lower electron recoil energy. At present, there are attempts to push $E_e^{\rm min}$ down to energies of 100\,MeV to 200\,MeV. However, this would typically imply a reduction in the reconstruction efficiency of the recoil electron and thus may not result in a significant increase in sensitivity.
In the near future the low mass region $ m_{\chi} \lesssim 60 $\,MeV may also be probed effectively by the \linebreak COHERENT experiment \cite{Akimov:2015nza,deNiverville:2015mwa,Ge:2017mcq}, whose primary purpose is measuring coherent elastic neutrino-nucleus scattering. 

\section{Conclusions}
\label{sec:conc}
We studied DM-electron scattering signatures in the NO$\nu$A  near detector.
Our main finding is that in the 6 MeV-200 MeV dark matter mass window NO$\nu$A can probe unexplored regions of the parameter space reaching down to the thermal relic line for some values of $\alpha_D$ for a complex scalar DM candidate. This represents a significant improvement over existing scattering experiments like LSND  \cite{ lsnde,lsndpatrick}  and E137 \cite{Batell:2014mga}.
Our proposal is completely symbiotic to the neutrino program and requires only a recast of an existing analysis. As a further step, a dedicated analysis would be welcome and could further improve the sensitivity. We have thus considered some preliminary ideas to improve the sensitivity.
As shown in Figs.\ref{escattering1} and \ref{escattering2}, the region of the parameter space within NO$\nu$A's reach exhibits a nice complementary with many planned future facilities and proposed experiments such as Belle-II \cite{Yuan:2012jd} and BDX \cite{Battaglieri:2016ggd}. In particular, if a discovery were made at NO$\nu$A it would then also be possible to study the properties of this particle. In particular, future direct detection experiments such as SENSEI \cite{cosmicvision,Crisler:2018gci} would be able to test if the discovered particle is DM (at least in the case of a complex scalar) since NO$\nu$A, as with other accelerator-based probes, cannot distinguish DM from other long-lived particles.

We conclude by commenting on the complementary of our proposal to that put forward in \cite{Coloma:2015pih,Frugiuele:2017zvx} which illustrated the possibility to probe few GeV leptophobic mediators between the visible and the dark sector using MiniBooNE data from neutrinos coming from the Main Injector.
Hence building a comprehensive DM program at Fermilab NUMI facility is possible and highly motivated.
\section*{Acknowledgments}
We would like to thank Alec Habig, Athans Hatzikoutelis and Biao Wang for valuable discussions regarding NO$\nu$A near detector.

\bibliography{NOVA}

\end{document}